\documentclass[runningheads]{llncs}

\usepackage[pdftex]{graphicx}

\usepackage{enumerate}
\def\circledenumerate#1{\textcircled{\scriptsize #1}}

\usepackage{comment}

\usepackage{amssymb,amsmath}
\usepackage{cancel}
\usepackage{url}

\newcommand{\ctext}[1]{\raise0.1ex\hbox{\textcircled{\scriptsize{#1}}}}
\newcommand{\maru}[1]{\raise0.2ex\hbox{\textcircled{\scriptsize{#1}}}}

\newcommand{\Any}{\texttt{\char`_}}

\begin{document}

\pagenumbering{arabic}
\pagestyle{plain}

\setlength\abovecaptionskip{0pt}
\setlength\textfloatsep{10pt}

\title{Coordination-free Collaborative Replication based on Operational Transformation}
\titlerunning{Coordination-free Collaborative Replication}

\author{Masato Takeichi}
\authorrunning{M. Takeichi}
\institute{University of Tokyo\\
 \email{takeichi@acm.org}
}

\date{\today}

\maketitle

\begin{abstract}

We introduce Coordination-free Collaborative Replication (CCR), a new method for 
maintaining consistency across replicas in distributed systems without requiring explicit 
coordination messages. CCR automates conflict resolution, contrasting with traditional 
data sharing systems that typically involve centralized update management or predefined 
consistency rules.

Operational Transformation (OT), commonly used in collaborative editing, ensures consistency by transforming 
operations while maintaining document integrity across replicas. However, OT assumes 
server-based coordination, which is unsuitable for modern, decentralized Peer-to-Peer 
(P2P) systems.

Conflict-free Replicated Data Type (CRDT), like Two-Phase Sets (2P-Sets), guarantees eventual consistency by allowing 
commutative and associative operations but often result in counterintuitive behaviors, such 
as failing to re-add an item to a shopping cart once removed.

In contrast, CCR employs a more intuitive approach to replication. It allows for 
straightforward updates and conflict resolution based on the current data state, enhancing 
clarity and usability compared to CRDTs. Furthermore, CCR addresses inefficiencies in 
messaging by developing a versatile protocol based on data stream confluence, thus 
providing a more efficient and practical solution for collaborative data sharing in distributed 
systems.

(Original Version submitted September 16, 2024)
\end{abstract}

\section{Introduction}
\label{sec:Introduction}

We propose a novel idea of \textit{Coordination-free Collaborative Replication} (CCR)
that guarantees the consistency of replicas under individual updates in
distributed systems. It makes automatic conflict resolution without
any explicit message exchange for coordination.

Operational Transformation~\cite{ellis1989concurrency,sun1998operational,sun2008context}
is a technology to guarantee consistency
when updating common data in collaborative data sharing.
For example, we put a replica of a text document at each site for editing.
Each site sends the editing operations done on the local replica to the server.
The server transmits operations to other clients to keep all the replicas
the same in distributed environments.
Each site appropriately applies the editing operations sent from the server
to its local replica.
For the transformation of text editing, the basic
operations of inserting and deleting character strings and moving
the cursor keeps the consistency that all
replicated documents are the same when the editing operation is
completed.
Various practical document processing
systems have been developed and used in this way.
A conflict may occur when character strings are inserted concurrently
to the same point at multiple sites.
To resolve this kind of conflict, we may take one of the insertions
results are made valid, or both are made effective by shifting either one.

While this approach is extendable to other structured documents,
OT-based data sharing usually assumes that the server manages updates
of clients.
However, this is not the case for recent applications of distributed systems.
We should be more generous and let ourselves to
make our replication to the server-less configuration for
Peer-to-Peer applications suitable for the \textit{local-first}
applications~\cite{kleppmann2019local}.
We explore the feasibility of defining OT for such systems and propose
a \textit{Collaborative Replication} technique.

CRDT~\cite{shapiro2011conflict,shapiro2011comprehensive,zhao2018observable}
is another method for ensuring consistency when data is
updated by an operation on the replica placed at sites.
The goal of the CRDT is to ensure eventual consistency,
where the replicas converge when updates terminate.
For that purpose, we make data closed
under commutative and associative operations.
There, the operations on the data are limited to satisfy the monotonic condition
which is not always clear from our intuitive aims of updating.

A shopping cart implemented using Two-Phase Set (2P-Set)
CRDT~\cite{shapiro2011comprehensive} is an example.
It accepts operations $({\sf{Add}} ~x)$ to add item $x$ into the cart and
$({\sf{Rem}} ~x)$ to remove item $x$ from the cart.
2P-Set consists of a pair $(A, R)$ that keeps added items
in $A$ and removed items in $R$, which grow every time these operations
are performed as $A \cup \{ x \}$ and $R \cup \{ x \}$. 
The items in the cart can be calculated as a set difference  $A \setminus R$.
This implementation fails the item $x$ added again when $x$ has once been added
and then removed.
Can this shopping cart be acceptable, or is this implementation intuitive?

In contrast to this CRDT shopping cart, Collaborative Replication implements
the cart $C$ to which $x$ is added as
$C \cup \{ x \}$ if $x \notin C$,
and from which $x$ is removed as $C \setminus \{ x \}$ if $x \in C$.
It is straightforward to specify how the cart $C$ is updated.
It enables us to add $x$ once having been removed.
Note here that these operations refer to the current status of $C$ for
checking the effectiveness of updates on $C$
called
\textit{Set with Effectful Operations}~\cite{takeichi2021conflict}.
It is much more understandable than CRDT implementation.

Another point to note in updating common data in a distributed environment
is how to eliminate inefficiency
due to messaging in the distributed system.
We have to study more about protocols for avoiding such inefficiency
as well as resolving conflicts related to updating.
We develop a concise but versatile protocol
based on the property of the \textit{confluence} of data stream updated
by operations.
It might be what we come to think of if we know it.

Combined these, we call this novel approach \textit{Coordination-free} protocol for
\textit{Collaborative Replication}.

\section{Collaborative Replication}
\label{sec:CR}

The problem of conflicts caused by concurrent updates at different sites
can be solved by either making adjustments in advance
so that no concurrent updates start, or by choosing one of the concurrent
updates.

The elimination of conflicting data updates involves a temporary 
shutdown of other sites other than the one updating.
When choosing one of the concurrent updates,
a slight difference in the update timing may cause
consistency problems.

In contrast, there are proposals to set certain restrictions on the nature of
the data to be shared, i.e., the operations to update the data, so that conflicts
themselves cannot occur or can be resolved to make the result in a deterministic way.
CRDT (Conflict-free Replicated Data Type) and
CCR (Coordination-free Collaborative Replication) aims in this direction.

\subsubsection{Shattering the CALM for Collaborative Data Sharing}
CRDTs and CCR are within the same framework of
collaboration in how
updates performed and how they reflected as shared data.
CCR has
aspects that are more concise and versatile than the structure of CRDTs,
and that is the focus of this discussion.

While coordination is a ``killer'' of performance in distributed systems, no
coordination may suffer from the consistency of distributed data.
The CALM (Consistency As Logical Monotonicity) theorem brings about a
solution to the question~\cite{hellerstein2020keeping}:
``What is the family of problems that can be consistently computed in a
distributed fashion without coordination, and what lies outside that family?''
as
``A program has a consistent, coordination-free distributed implementation if
and only if it is monotonic.''

CALM leverages static analysis to certify the state-based convergence properties
provided by CRDTs which provide a framework for monotonic programming
patterns.

However, monotonicity is not the only golden rule for coordination-free data sharing in
distributed systems!
A binary operation is commutative if the order of its operands makes no
difference to the result.
The confluence property
as applied to program operations is a generalization of commutativity.

\begin{itemize}
\item An operation is confluent if it produces the same outputs for any
nondeterministic ordering of a set of inputs.
\item If the outputs of one confluent operation are consumed by another confluent
operator as inputs, the resulting composite operation is confluent.
\item Hence, if we restrict ourselves to building programs by composing confluent
operations, our programs are confluent by construction, despite orderings of
messages or execution steps within and across sites of distributed systems.
\end{itemize}
Although confluent operations are the basic constructs of monotonic systems,
they can do more than that if collaboration is utilized for establishing confluence
of components of distributed systems.

Thus, we can shatter the restricted CALM for monotonicity to open up more
flexible CCR than CRDTs.

\subsubsection{Operational Transformation is Disliked in Replication, but ...}
Compared with the CRDT approach, Operational Transformation originally proposed
in 1989 has been kept away from replicated data in server-less systems since most algorithms were proved to be
wrong~\cite{sun2020real,laddad2022keep}.

In OT, every updating process is a sequence of a patch of operations 
transmitted between sites. Received operations in each site are transformed to
include the local operations performed since the last common data.
They are then applied locally to get the new data.

For high-level operations like server-less text editing
such transformation has too many edge cases
which are difficult to produce in the same state.
However, it is not in many cases of selected operations dealt with by
CRDTs.
Or more than that, we should notice that OT generates ``new operations''
from the already performed local operations for making the states confluent
and hence leads to more flexible ``collaborative'' replication.
It is in contrast with CRDT where the predefined
common operations are used in both sites.

Our CCR is a new challenge to exploit OT-based data sharing.
It comes from the Tutorial Blog\footnote{{http://blog.haskell-exists.com/yuras/posts/realtime-collaborative-editor.html}}
 ``Realtime collaborative editor. Algebraic properties of the problem''
by Yuras Shumovich,
March 5, 2016
which explains concisely the idea and pitfalls of Collaborative Editor (RTCE) implementations
based on operational transformation.
Although it assumes that the system uses a server for maintaining the state of clients,
it works in a server-less environment, too.

OT-based Collaborative Replication is very lightweight without extra metadata essential in CRDTs,
and hence enable composing replicated data from simple data as the popular
abstract data definitions using data constructors, e.g., tuples, maps, etc.
It is what we claim our CCR to exploit
a novel scheme for consistent data sharing.


\section{Collaborative Replication based on Operational Transformation}
\label{sec:CR-OT}

\subsection{Updating Operations and Operational Transformation}

\subsubsection{Updating Operations}
Updating operations on the replica are defined for manipulating the replica to share the data according to the user's intention.
\begin{itemize}
\item The set $O$ of operations on data $D \in \mathcal{D}$ is
a \textit{monoid} $(O,!,\odot)$ with the identity operation ``$!$'' with
several basic operations as generators and $\odot :: O \times O \rightarrow O$.
\item Updating operator $p \in O$ applied to $D$ is generically
written as $D \odot p$ using a binary postfix operator symbol
$\odot :: \mathcal{D} \times O \rightarrow \mathcal{D}$,
which is overloaded with the monoid operation.
Thus, $D \odot p \odot q = (D \odot p)\odot q = D \odot (p\odot q)$.
\item The sequence of updating operators $p_1, p_2, \cdots$ is
called \textit{patch} and written as $\langle p_1, p_2, \cdots \rangle$,
which is of course an operator in $O$.
\item The equality on $\mathcal{D}$ of data $D \odot p$ and $D \odot q$
both of which are $D$ applied updating operators $p$ and $q$ respectively
defines the equivalence relation $p \leftrightarrow_D q$ on $O$.
This defines the equality of operators $p=_D q$.
Note that these relations depend on $D$ to which the operators are applied.
\end{itemize}

In practical situations, basic operations $p$ and $q$
are defined on the data $D \in \mathcal{D}$ to be replicated.
For example, basic operators for text editing are $({\sf{Ins}}~k~ t)\in O$ for inserting
string $t$ at position $k$ into the replicated text,
and $({\sf{Del}} ~k~n) \in O$ for deleting $n$ characters at $k$ from the replica.
These operations are usually ``generators'' of the monoid applied
to the replicated data.
Successive application of these generator operators to the data can be
expressed using the application of the patch of these operators, e.g.,
$\langle {\sf{Ins}}~k_1~ t_1, {\sf{Del}} ~k_2~n_2, {\sf{Ins}}~k_3~ t_3\rangle$
which is a representation of the element
$({\sf{Ins}}~k_1~ t_1 \odot {\sf{Del}} ~k_2~n_2 \odot {\sf{Ins}}~k_3~ t_3) \in O$
of the monoid $(O,!,\odot)$. Note that the element of the monoid may have
several patch representations.

\subsubsection{Operational Transformation}

Operational Transformation $T_D :: O \times O \rightarrow O \times O$ takes operations $p$ and $q$ on $D$ and gives a pair of operations $p'$ and $q'$ by $(p',q')=T_D (p,q)$.

\setlength\abovecaptionskip{-10pt}
\begin{figure}[htb]
 \begin{center}
  \includegraphics
  [width=0.7\textwidth]{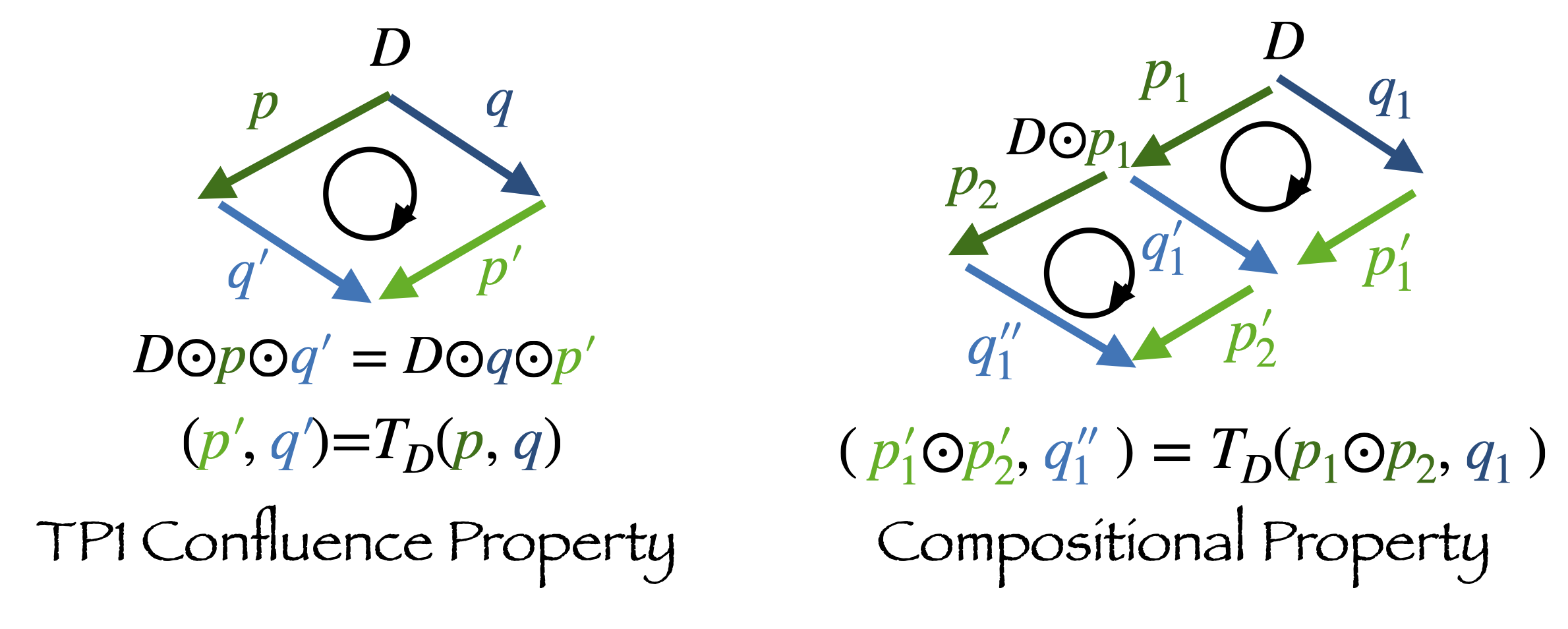}
 \end{center}
  \caption{Operational Transformation for Replication}
  \label{fig:OT}
\end{figure}

Our OT for replication is assumed to have the following properties:

\begin{description}

\item{\bf{Local Confluence Property}} The transformation $T_D$
takes operations $p$ and $q$ on $D$ and produces $(p',q')=T_D (p,q)$
which makes $D \odot p \odot q' = D \odot q \odot p'$.
That is, $p$ and $q$ on the same $D$ are made into the common confluent state
by $T_D$. This property is called ``TP1-Confluence'' which is the basis of
``TP2-Confluence'' for
global integration stability described below.

\item{\bf{Minimal Property}} If $(p',q')=T_D (p,q)$ holds,
there is no operations $p''$, $q''$, $r_p \ne !$ and $r_q \ne !$ satisfying
$(p'', q'')=T_D(p,q)$ where $p' = p'' \odot r_p$ and $q' = q'' \odot r_q$.
That is, no extra operations are introduced by $T_D$.

\item{\bf{Symmetric Property}} If $(p',q')=T_D (p,q)$ and $(q'',p'')=T_D (q,p)$,
then $p'=p''$ and $q'=q''$ hold.

\item{\bf{Compositional Property}} If $(p_1',q_1')=T_D (p_1,q_1)$ and
$(p_2',q_1'')=T_{D\odot p_1} (p_2,q_1')$, $T_D (p_1 \odot p_2,q_1)$
gives $(p_1' \odot p_2', q_1'')$. 

\end{description}

These properties are the most fundamental and more are required for advanced
purposes as described in the next section.
Before going there, we review the presumption of this property.
\begin{itemize}
\item In case that $p$ and $q$ have their inverse $p^{-1}$ and $q^{-1}$, that
is the operation monoid $(O,!,\odot)$ is a semigroup, we may define $T_D$
as $T_D(p,q)=(q^{-1},p^{-1})$ which always makes $D$ into the confluent data
$D$. However, no one would like to use this transformation.

\item The Symmetric Property is natural for server-less P2P replication where
there is no dedicated site for the server to control the update propagation and
all the sites in the system are on an equal footing with others.
One of the transformations for text editing claimed to be correct is actually
non-symmetric because it is aimed solely at the environment of the server-clients
system.
\item The Compositional Property is very natural for the transformation is
defined for the basic operations and extends it to accept any operations of the
monoid $(O,!,\odot)$.
\end{itemize}

\subsection{Building up Confluence Property of Operational Transformation}
\label{subsec:Confluence}

Operational transformation $T_D :: O \times O \rightarrow O \times O$ for two
sites $P$ and $Q$ gives $(p_q,q_p)=T_D (p,q)$
for operations $p$ and $q$ respectively,
and fulfills \textit{TP1-Confluence} $p \odot q_p \leftrightarrow_D q \odot p_q$,
which we write as $p \#_D q$.
When $D$ is clear from the context, $\#_D$ may be written simply as $\#$.

\subsubsection{Idempotence of Confluence Relation}
When $P$ and $Q$ share a common updating operation $p$ applied to the
same data $D$ to get $D \odot p$, it is natural to assume that the confluent
state is $D \odot p$ and hence $p \#_D p = p$.

Although this conforms to the Minimality of $T_D$,
it is subtle to define $T_D(p, p)=(!, !)$ that implies $p \#_D p = p$.
As mentioned in
Section~\ref{sec:Introduction} in the text editing application,
a conflict may occur when strings $x$ and $y$ are inserted
to the same point $k$ by operations $p=({\sf{Ins}}~k ~x)$
and $q=({\sf{Ins}}~k ~y)$.
To resolve the conflict,
we may reasonably make both insertions effective by shifting $q$'s
$k$ to $k'=k+ length(x)$ to define
$T_D({\sf{Ins}}~k ~x, {\sf{Ins}}~k ~y)=({\sf{Ins}}~k ~x, {\sf{Ins}}~k' ~y)$.
However, if $x$ and $y$ are equal, this contradicts $T_D(p, p)=(!, !)$
and $p\#_D p=p$.

What happens there?
So far, we have never been concerned about the independence of the
operations applied in different sites:
above $p$ and $q$ look the same for $x=y$ in spite of their independence.
Hereafter to make the independence explicit by 
attaching the site ID $\$i$ in the operation as $({\sf{Ins}} ~\$i~k ~x)$
for the internal representation of the operation, while the external
representation remains as $({\sf{Ins}} ~k ~x)$.

\subsubsection{Commutativity of Confluence Relation}
It is straightforward that the confluent relation $\#_D$ is commutative, that is,
$p \#_D q = q \#_D p$ for operations $p$ and $q$ on the same data $D$.

\subsubsection{Associativity of Confluence Relation}
Then, how about the confluence relation for three sites $P$, $Q$, and $R$?
Consider updates by three sites implemented by applying the transformation
$T_D$ in two steps.
We first apply replication of $P$ with another site $Q$ or $R$, and then apply
next with the rest of $R$ or $Q$.
Thus,

\begin{itemize}
\item{$P \rightarrow Q \rightarrow R$:}
$P$ having updated $D$ into $D \odot p$ receives $Q$'s update $q$ and computes
$(q_p, \Any)=T_D(q,p)$ to produce $D \odot p \odot q_p$, and then 
receives $R$'s update $r$ and computes
$(r_{p \odot q_p}, \Any)=T_D(r, p \odot q_p)$
to get $D \odot (p \odot q_p)\odot r_{p \odot q_p} = D \odot ((p \#_D q) \#_D r)$.
\item{$P \rightarrow R \rightarrow Q$:}
$P$ having updated $D$ into $D \odot p$ receives $R$'s update $r$ and computes
$(r_p,\Any)=T_D(r,p)$ to produce $D \odot p \odot r_p$, and then 
receives $Q$'s update $q$ and computes
$(q_{p \odot r_p}, \Any)=T_D(q, p \odot r_p)$
to get $D \odot (p \odot r_p)\odot q_{p \odot r_p} = D \odot ((p \#_D r) \#_D q)$.
\end{itemize}

Here, we could say that it is necessary to assure
$(p \#_D q) \#_D r = (p \#_D r) \#_D q$
for each site to lead its data to the confluent state.
Similarly, other cases of permutations of $P$, $Q$, and $R$,
namely $Q \rightarrow R \rightarrow P$, $Q \rightarrow P \rightarrow R$,
$R \rightarrow P \rightarrow Q$, and $R \rightarrow Q \rightarrow P$.

To establish the confluence property of $T_D$ applied in two steps for updates
on the same data $D$ by three sites,
the pairwise equivalent relations need to hold regardless of
the order of applications. That is,
$(p \#_D q) \#_D r = (p \#_D r) \#_D q$,
$(q \#_D r) \#_D p = (q \#_D p) \#_D r$, and
$(r \#_D p) \#_D q = (r \#_D q) \#_D p$
hold.

\setlength\abovecaptionskip{-10pt}
\begin{figure}[htb]
 \begin{center}
  \includegraphics
  [width=0.5\textwidth]{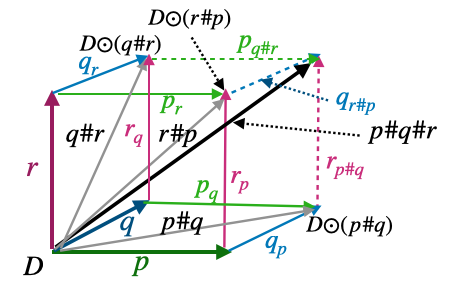}
 \end{center}
  \caption{TP2 Confluence Property}
  \label{fig:TP2}
\end{figure}

Therefore, we assume that our OT for collaborative replication must satisfy
the associativity property of the confluence relation.
That is, for operations $p$, $q$ and $r$ on the same data,
\[ (p \#_D q) \#_D r = p \#_D (q \#_D r) \]
holds, and these may be written as $ p \#_D q \#_D r$ without parentheses.

\subsubsection{TP2-Confluence Property}

The above confluence property is a case of
\textit{TP2-Confluence Property} widely known
in the OT community~\cite{oster2005proving,randolph2014synthesizing,gomes2017verifying}.
This requires a confluence of the transformation for more than three
sites in general. That is, it should hold for more than four, not just for three.

However, it is ready to extend the above story to the cases of more than four
sites.

\setlength\abovecaptionskip{-10pt}
\begin{figure}[htb]
 \begin{center}
  \includegraphics
  [width=0.7\textwidth]{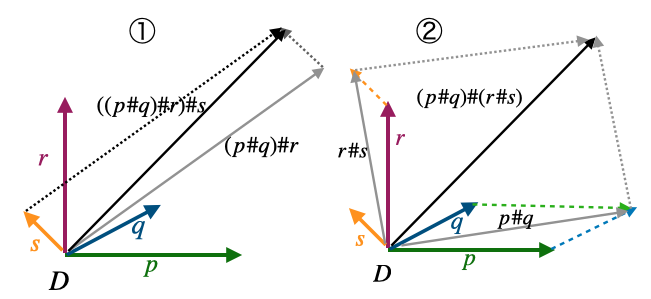}
 \end{center}
  \caption{Confluence of Four Sites}
  \label{fig:FourSites}
\end{figure}

Updates of four sites are implemented by three applications of $T_D$;
the last step of applications may take a fresh operation $s$ with the confluent
operation $p \#_D q \#_D r$ of the three, or take the confluent operations $p \#_D q$
and $r \#_D s$ of the two.
The confluence property of two-step applications up to three sites is assured
as above.
Therefore, updates of four sites are made confluent.
This inductively
concludes that the associativity and commutativity properties of the
confluence relation assure the general confluence of any number of sites.
This is what the TP2-Confluence Property holds.

\subsubsection{Algebraic Structure of Confluence Property}
\begin{description}
 \item{\bf{Identity element}} $p \#_D ! = ! \#_D p = p$
 \item{\bf{Idempotence}} $p \#_D p = p$, which comes from
 the Minimal Property of $T_D$
 \item{\bf{Commutativity}} $p \#_D q = q \#_D p$,
 which comes from TP1-Property of $T_D$
 \item{\bf{Associativity}} $(p \#_D q) \#_D r = p \#_D (q \#_D r)$,
 which is for TP2-Property of $T_D$
\end{description}

\setlength\abovecaptionskip{-10pt}
\begin{figure}[htb]
 \begin{center}
  \includegraphics
  [width=0.9\textwidth]{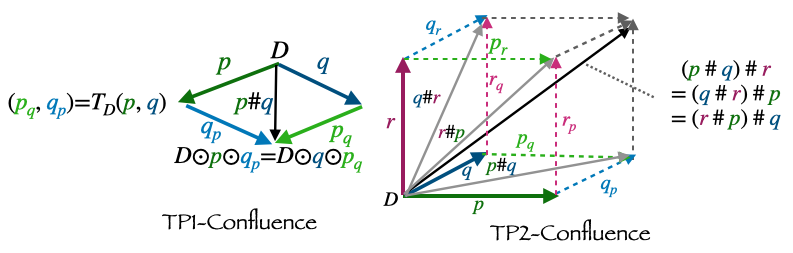}
 \end{center}
  \caption{TP1+TP2 Confluence Property}
  \label{fig:TP1TP2}
\end{figure}

The above discussion suggests us to put the TP2-Confluence Property
into practical coordination-free replication by sending updated operations
on the common replicated data in arbitrary order to others.

\section{Replication Protocol for Coordination-free Collaborative Replication}
\label{sec:CCR}

\subsubsection{Replication Protocol of Operation-based CRDT}
In Operation-based CRDT (CmRDT), the exchange of updated operations
for replication is performed by
Reliable Causal Broadcast (RCB), while in State-based CRDT (CvRDT),
by merging data states (merge).

RCB for exchanging updates
must realize that every local update at each site is eventually reflected
at other sites.
Local updates at a site
must be sent to the other site again in the event of a disconnection or
system outage after a local update.
According to the CmRDT definition, when a site sends a message to other sites, there is no confirmation of its arrival by the other sites,
and each site can operate even when disconnected.

\subsubsection{Replication Protocol of OT-based Collaboration}
Although OT-based replication shares some similarities
with CmRDT, its protocol does not necessarily follow RCB.

OT-based collaborative data sharing also realizes reliable replication by broadcast,
where all the local updates eventually arrive on all sites.
The causality of operation events is not built into the protocol,
but it is handled in the definition of OT with
information accompanying the update operation.
In OT-based replication, each site receives a sequence of operations to be
applied to each site by converting a sequence of operations (patch),
and each site applies them in a cooperative manner (collaboration) to achieve
replication

In CmRDT, operations are defined independently from the data state, but in
OT-based replication, operations are defined on the data state.
Unlike CmRDT operations, update operations in OT-based replication are not
necessarily commutative, so the result of individual replication may depend on
the order in which operations are applied.
The OT's compositionality guarantees that replication is possible for three
or more sites by applying a combination of OTs that define a two-party replication,
and also guarantees the Coordination-free Collaborative Replication.

\subsubsection{Asynchronous Messaging for Collaborative Replication\\}


Collaborative replication in distributed sites can be possible by exchange
messages of their updated information. A simple way to do this would be like this:
\begin{itemize}
\item When sites $P$ and $Q$ perform
operations $p$ and $q$ independently and concurrently to get their states
$D \odot p$ and $D \odot q$ respectively,  computing $(p_q, q_p)= T_D (p,q)$
from $p$ and $q$ brings both states into the confluent state
$D\odot p \odot q_p = D\odot q \odot p_q$.
 \item But this is not  practical because $P$ and $Q$ cannot share both $p$ and $q$
at the same time.
Either of $P$ or $Q$, say $P$ first sends $p$ to $Q$ and then $Q$ receives it
and computes $p_q$ by $(p_q, \Any)= T_D (p,q)$ to update the state into
$D \odot q \odot p_q$.
\item For $P$ to update its state to $D \odot p \odot q_p$,
it requires $Q$ to send $q$ just when $Q$ receives $p$ from $P$.
If no other sites exist, $P$ merely waits for $Q$'s reply of $q$.
Otherwise, for $P$ to get $Q$'s reply without the intervention of other's updates,
$P$ and $Q$ need to keep step update exchange between them.
\item To do this, $P$ blocks communication with other sites except $Q$
until it receives $Q$'s reply.
\end{itemize}

Such control to synchronize messages to wait messages of the particular site
needs to spread blocking commands globally across the distributed sites of
the system. It causes a great loss of efficiency and needs complex
concurrency control.

On the contrary, our CCR scheme is simple enough to send an update
information at any time
without any synchronizing control, that is, asynchronous messaging in a way
that each site
sends updating information at any time and receives information from
any site at any time.
It relies on the algebraic structure of the confluence property of our
Operational Transformation.

Our coordination-free asynchronous replication proceeds as follows:

\setlength\abovecaptionskip{-10pt}
\begin{figure}[htb]
 \begin{center}
  \includegraphics
  [width=0.9\textwidth]{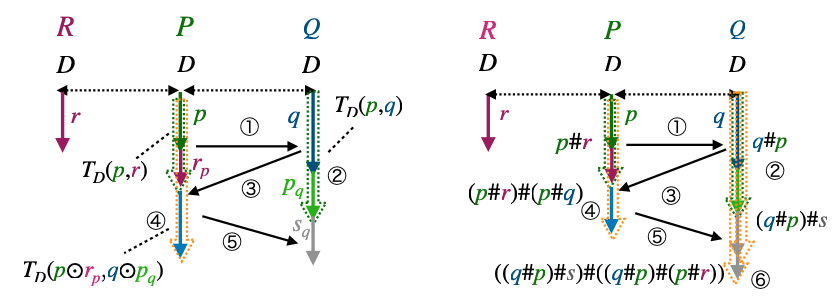}
 \end{center}
  \caption{Asynchronous Messaging in Collaborative Replication}
  \label{fig:Asynchronous}
\end{figure}

\begin{enumerate}[\expandafter \circledenumerate 1]
\item $P$ sends $p$ to $Q$ and other sites at its convenient time.
\item $Q$ computes $(p_q,q_p)=T_D(p,q)$ to get the operation $q \odot p_q$
that makes $D$ into the confluent state $D\odot q \odot p_q$.
Note that $q \odot p_q$ is a representation of $q \#_D p= p \#_D q$
which stands for $q \odot p_q \leftrightarrow_D p \odot q_p$ determined by
$T_D(p,q)$. $Q$ keeps $q \odot p_q$ for its local representation of $q \#_D p$.
\item Then, $Q$ sends $q \#_D p$ to other sites including $P$. Note that $Q$ 
actually broadcasts the representation $q \odot p_q$ of the operation $q \#_D p$.
\item When $P$ receives $q \odot p_q$ sent from $Q$ after its state $D$
has been updated to $D \odot p \odot r_p$ by some $r$ from $R$,
$P$ makes $p \odot r_p$ and $q\odot p_q$ confluent into
$(p \odot r_p)\#_D (q \odot p_q)$ by computing $T_D(p \odot r_p, q \odot p_q)$.
\item $P$ broadcasts this to other sites. Here, $p \odot r_p$ is a representation
of $p \#_D r$ and $q \odot p_q$ is a representation of $q \#_D p = p \#_D q$,
and then $(p \odot r_p)\#_D (q \odot p_q)$ is equivalent to
$(p \#_D r)\#_D (p \#_D q)= p \#_D (r \#_D q)$.
\item In the same manner, when $Q$ has reflected $S$'s update $s$
as $(q \#_D p)\#_D s$ and receives $(p \#_D r)\#_D (p \#_D q)$
and computes a representation of the new
operation confluent to $((q \#_D p)\#_D s)\#_D ((p \#_D r)\#_D (p \#_D q))$
which is equivalent to
$(q \#_D p)\#_D (s \#_D (p \#_D r))$.
This can be read as ``$Q$'s operation $q$ is first made confluent with $p$
and then with $s \#_D (p \#_D r)$,
which has been made confluent with $s$ and $p \#_D r$, and so on.
\end{enumerate}

We can manipulate the confluence relation according to the
algebraic property.
Note that the representation of the confluent operation $p\#_D q$ differs
site by site depending on the OT application.

\subsubsection{Eventual Consistency of Collaborative Replication}
As we have seen above, CCR assures that every update on the local
replica of a site is sent to other sites to produce the new confluent
operations in each site.
Every time updates occur in the system, they continue 
the replication process of making a new confluent state.
When no updates are in the system, the replicated data in the sites is the same
across the system.
Thus, the confluent replication leads to the eventual consistency.

But when does the replication process terminate?
If $R$ has not operated $r$ in step \ctext{4},
$P$ makes the confluent operation $p\#_D(q \odot p_q)$ or
$p \#_D (q \#_D p)$ which is reduced to $p \#_D q$ of which representaion
is $p \odot q_p$.
This implies that if $S$ has neither done $s$ in \ctext{6} then
$Q$ computes $T_D(q \odot p_q, p \odot q_p)$ which is $(!,!)$
by the compositional property.
Note that in this context, $p \odot q$ stands for the operation
sequence $\langle p, q \rangle$.
The above observation shows that the replication process should terminate
when OT produces
the identity operation.

Moreover, this implies that the update propagation over the circular
connection safely terminates. 

\subsubsection{Incremental Messaging in Collaborative Replication}
Up to now, we have not been concerned about the problem of efficiency about
the size of messages.
It costs too much if updated operations of each site are always sent
from the beginning of replication.
We can reduce the message size between connected sites
by keeping the indices of the last message of the partner site.

Note that the indices vary partner-wise of the connection.
It is an easy way of reducing the cost of messaging.

\section{Coordination-free Collaborative Replication in Practice}
\label{sec:CCRPractice}

\subsection{CCR Agent}
An experimental but being straightly put into practical use Agent
for Coordination-free Collaborative Replication is built for
confirming the principles and assumptions.
It is written in Haskell in several hundred lines of code using standard libraries.

The CCR Agent consists of modules, $Agent$, $AgentMain$, and $Replica$.
The module $Agent$ does Collaborative Replication for the replica defined in
the module $Replica$ with coordination-free asynchronous messaging with
other Agents. 
The $AgentMain$ module provides a REPL (Read-Eval-Print-Loop) for the user
asks the Agent to make or break WebSocket connections with other Agents and asks
to put update operations on the local replica.
These modules are independent of replica data type.

The $Replica$ module defines the type of the replica data, the operations on
the replica, and the transformation.
Some examples follow.

\setlength\abovecaptionskip{-10pt}
\begin{figure}[htb]
 \begin{center}
  \includegraphics
  [width=0.6\textwidth]{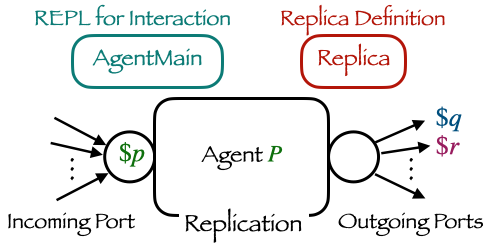}
 \end{center}
  \caption{Configuration of CCR Agent of 3 Modules} 
  \label{fig:CCRAgent}
\end{figure}

\subsection{Realtime Collaborative Editing}
\begin{figure}[htb]
 \begin{center}
  \includegraphics
  [width=1.0\textwidth]{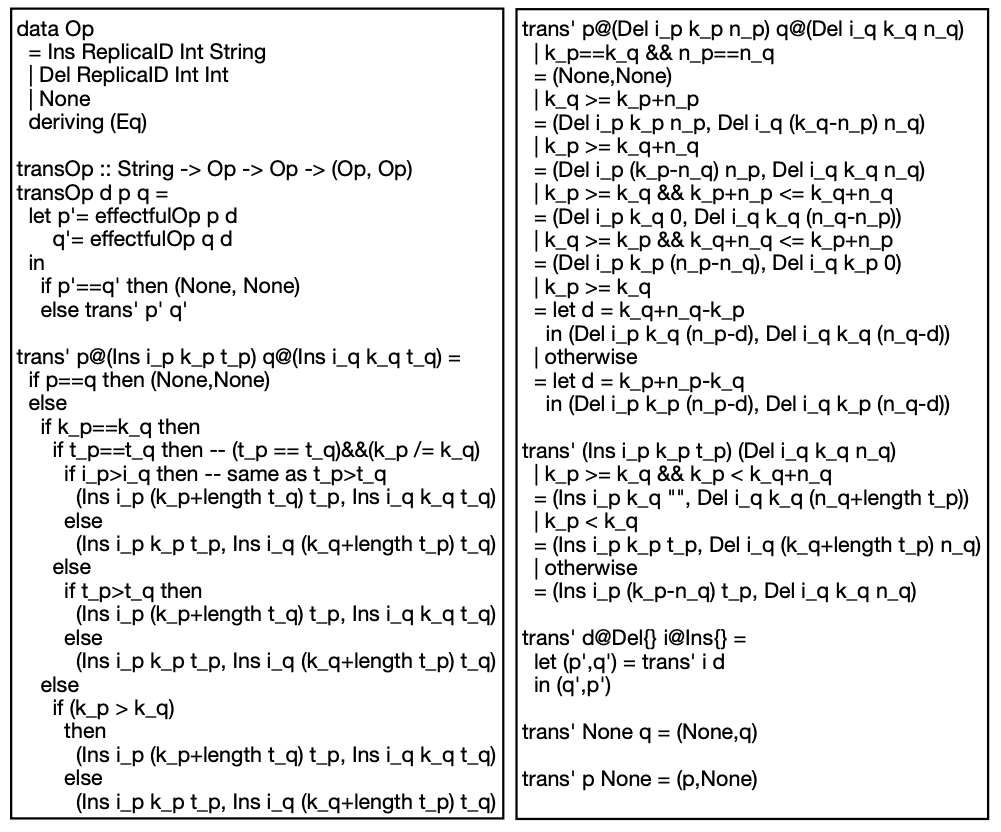}
 \end{center}
  \caption{Operational Transformation for Realtime Collaborative Editing}
  \label{fig:RTCE}
\end{figure}

As mentioned above RTCE pushed me into resolving to extend
OT-based data sharing to server-less collaborative replication.
An operational transformation satisfying our requirement is shown in Fig.\ref{fig:RTCE}.

While the algorithm should be verified by formal proofs, it has been tested by
more than a thousand combinations of randomly generated operations at three sites.
The formal proof is left to our further subject.

\subsection{Basic Replica Data}

\subsubsection{Counter}
Operations are $({\sf{Incr}}~\$i~n)$ for increment $D::Int$ by $n$ and
$({\sf{Decr}}~\$i~x)$ for decrement $D$ by $n$.
The parameter $\$i$ is the ID $(=0,1,2,\cdots)$ of the site in which
the operation is issued.
See the discussion on the Idempotence of Confluence in
Section~\ref{subsec:Confluence}.
The equality of operations is decided by comparing all the parameters, i.e.,
$\$i$ and $n$.

\begin{figure}[htb]
 \begin{center}
  \includegraphics
  [width=1.0\textwidth]{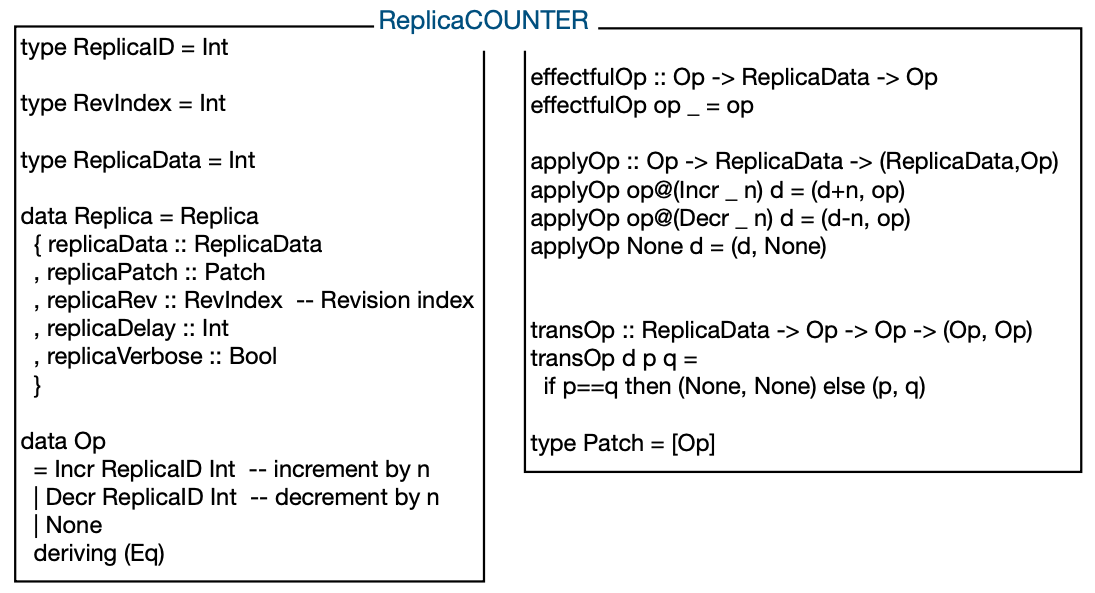}
 \end{center}
  \caption{Counter Replica}
  \label{fig:ReplicaCOUNTER}
\end{figure}

Definitions of the replica and the operations with transformation for the
$Int$ Counter
are in Fig.\ref{fig:ReplicaCOUNTER}.

The \textit{PN-Counter} CRDT implements the counter replica by collecting
local amounts of separated counters for increments and decrements.
It is much more elaborate than the CCR Counter.

\subsubsection{LWW Register}
The LWW (Last-Write-Win) Register is a typical CRDT that uses metadata
for maintaining the causality of events.
The updated data is associated with the timestamp of the event that is
compared when concurrent updates occur.
The LWW policy chooses the most recent one of these events
by comparing the timestamps.
While the timestamp may be the wall clock time, some logical clock
is preferred to avoid the clock skew problem.
However, the logical clock time is ordered in partial rather than total in general, and some
rules of conflict resolution are required.

In the demonstration of LWW\_String in CCR, the result strings by concurrent updates
are kept as a set of strings, each of which is the candidate of the winner. 

\begin{figure}[htb]
 \begin{center}
  \includegraphics
  [width=1.0\textwidth]{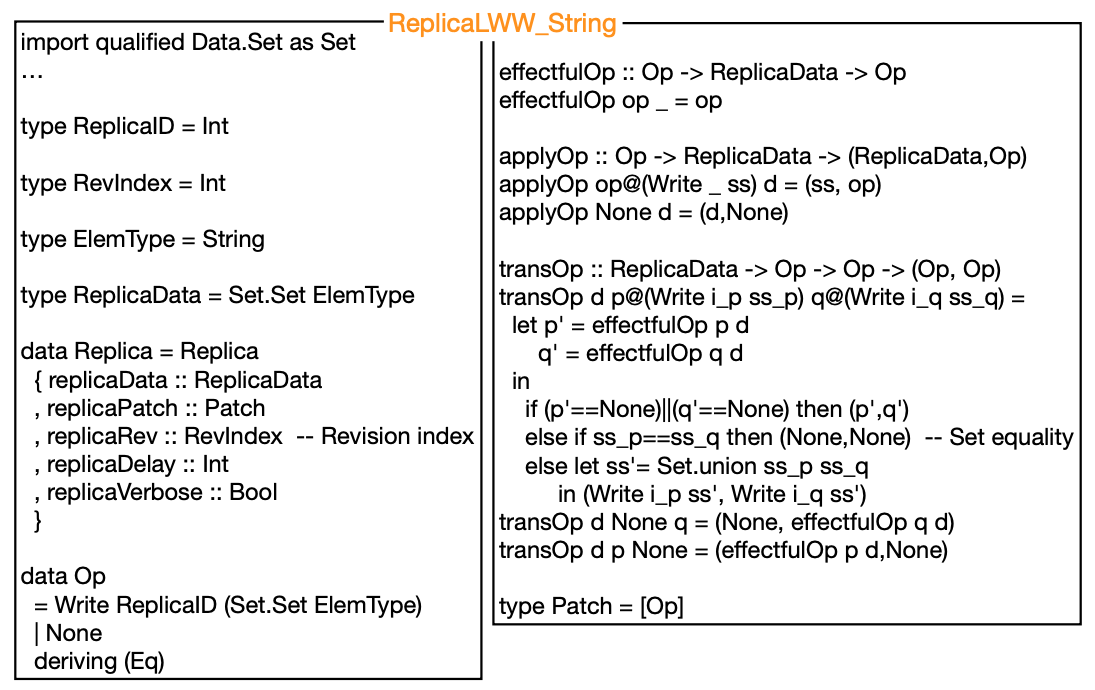}
 \end{center}
  \caption{Last-Write-Win Register of Strings}
  \label{fig:ReplicaLWW_String}
\end{figure}

Operations are $({\sf{Write}}~\$i~x)$ for replacing the set of string $D$ by  string $x$.
When there are no concurrent updates but one, $D$ becomes a singleton set.
Definitions of the replica and operations with transformations for an LWW of
strings are in Fig.\ref{fig:ReplicaLWW_String}.

\subsubsection{ESet}
In Section~\ref{sec:Introduction}, we have seen the shopping cart example
implemented by the set with effectful operations.
Operations are $({\sf{Add}}~\$i~x)$ for adding a element $x$ into the Set replica
$D$ if $x \notin D$ and $({\sf{Rem}}~\$i~x)$ for removing a element $x$ from
$D$ if $x \in D$.
The internal parameter $\$i$ is the ID $(=0,1,2,\cdots)$ of the site in which
the operation is issued.
The equality of operations is determined by comparing all the parameters, i.e.,
$\$i$ and $x$.

\begin{figure}[htb]
 \begin{center}
  \includegraphics
  [width=1.0\textwidth]{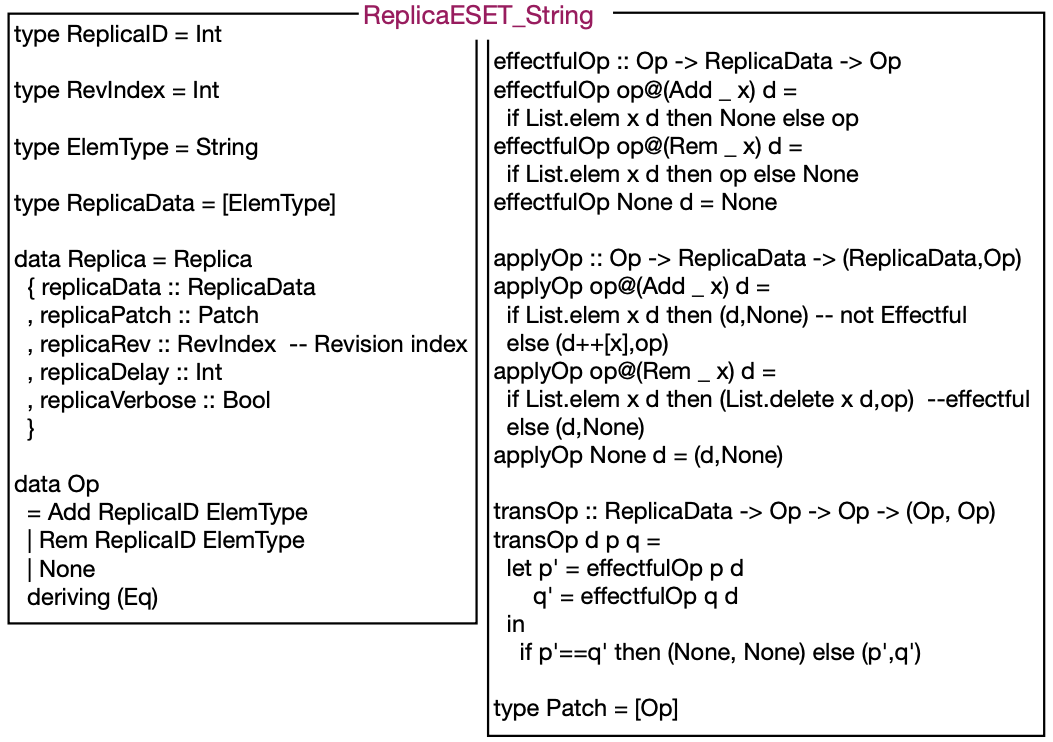}
 \end{center}
  \caption{Set of Strings with effectful operations}
  \label{fig:ReplicaESET_String}
\end{figure}

Definitions of the replica and operations with transformations for a Set of
strings are shown in Fig.\ref{fig:ReplicaESET_String}.

\subsubsection{Mergeable Replicated Queue}

Mergeable Replicated Queue is an example that demonstrates a framework
using the relational specification of the replicated data
\cite{kaki2019mergeable}.

Operations are $({\sf{Enq}}~\$i~x)$ for enqueuing an element $x$ into the queue replica
$D$ and $({\sf{Deq}}~\$i)$ for dequeuing an element $x$ from
$D$ if it exists.
The queue is represented by a pair of lists $(deq,enq)$ for the dequeued elements $deq$
and for the enqueued elements $enq$.
When concurrent enqueue operations occur, both elements are put into $enq$
in the same way as insertion operations at the same location in RTCE.

\begin{figure}[htb]
 \begin{center}
  \includegraphics
  [width=1.0\textwidth]{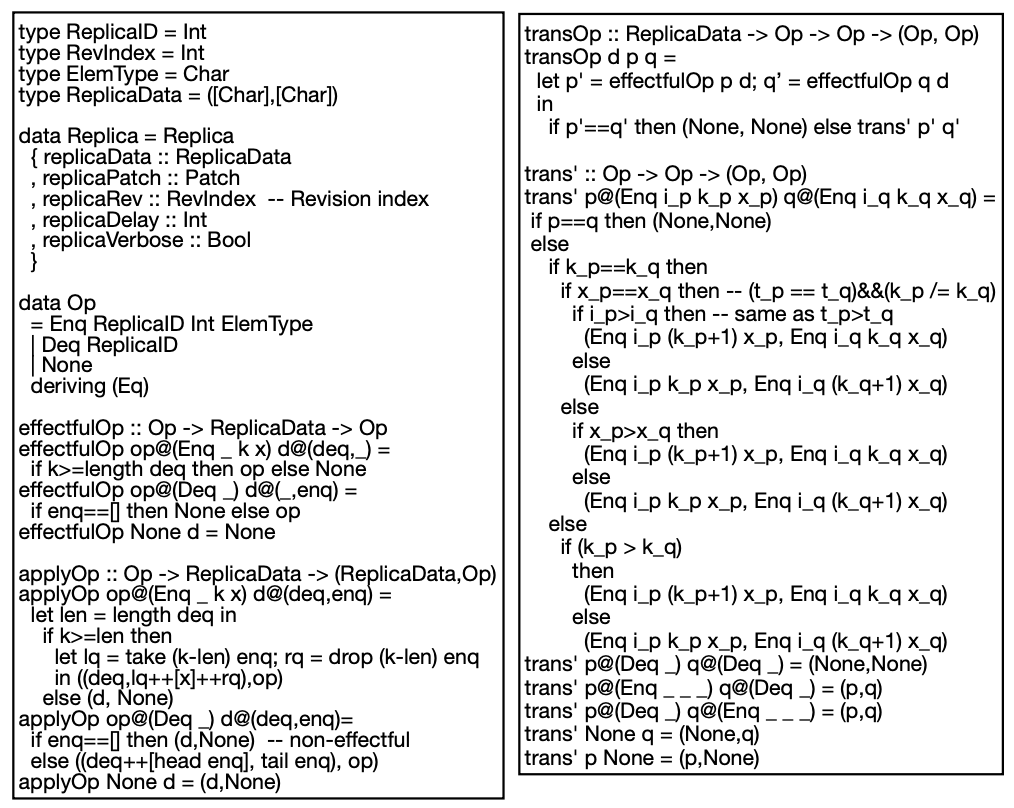}
 \end{center}
  \caption{Mergeable Replicated Queue}
  \label{fig:MRQ}
\end{figure}

Definitions of the replica and operations with transformations for a Set of
strings are in Fig.\ref{fig:MRQ}.

\subsubsection{Integer register supporting add and mult operators}
M. Weidner proposes \textit{Semidirect Product} for CRDT~\cite{weidner2020composing}
that makes concurrent updates by addition ${\sf{Add}}$ and multiplication ${\sf{Mult}}$ 
operations confluent. This rule supersedes the ordinal distributive law for
retaining the monotonicity of CRDT.

CCR transforms $({\sf{Add}}~m)$ and $({\sf{Mult}}~n)$ into
$({\sf{Add}}~m\times n, {\sf{Mult}}~n)$ to make $D$ become the confluent
$(D + m)\times n = (D \times n)+(m\times n)$,
which illustrates the semidirect product law using operational transformation.

\begin{figure}[htb]
 \begin{center}
  \includegraphics
  [width=1.0\textwidth]{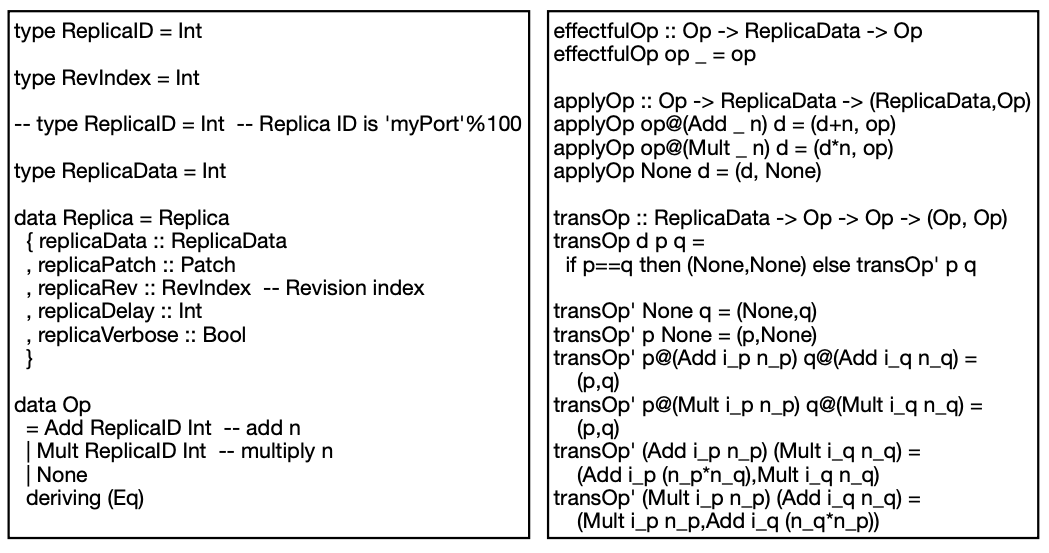}
 \end{center}
  \caption{Add-Mult Replication}
  \label{fig:AddMult}
\end{figure}

Definitions of the replica and operations with transformations for a Set of
strings are in Fig.\ref{fig:AddMult}.

\subsection{Composite Replica Data}
The construction of replicated replicas of composite data is a remarkable
feature of CCR.
It is as concise as the construction of composite data structures using
data constructors for tuples and maps.

\textit{Algebraic Replicated Data Types} proposed in~\cite{kuessner2023algebraic}
gives an example of a $SocialMedia$ replica for
a local-first social media application used by a group of friends in a P2P network
to share the quadruples $SocialPost$
of $messages$, $comments$, $likes$, and $dislikes$.
The $SocialMedia$ replica is the map of $SocialPost$ records.

While the ARDT approach assumes CRDTs with an intricate composition
of the replicated data
and operations,
CCR deals with this in constructive programming.

\subsubsection{Tuple of Replica Data}
\begin{figure}[htb]
 \begin{center}
  \includegraphics
  [width=1.0\textwidth]{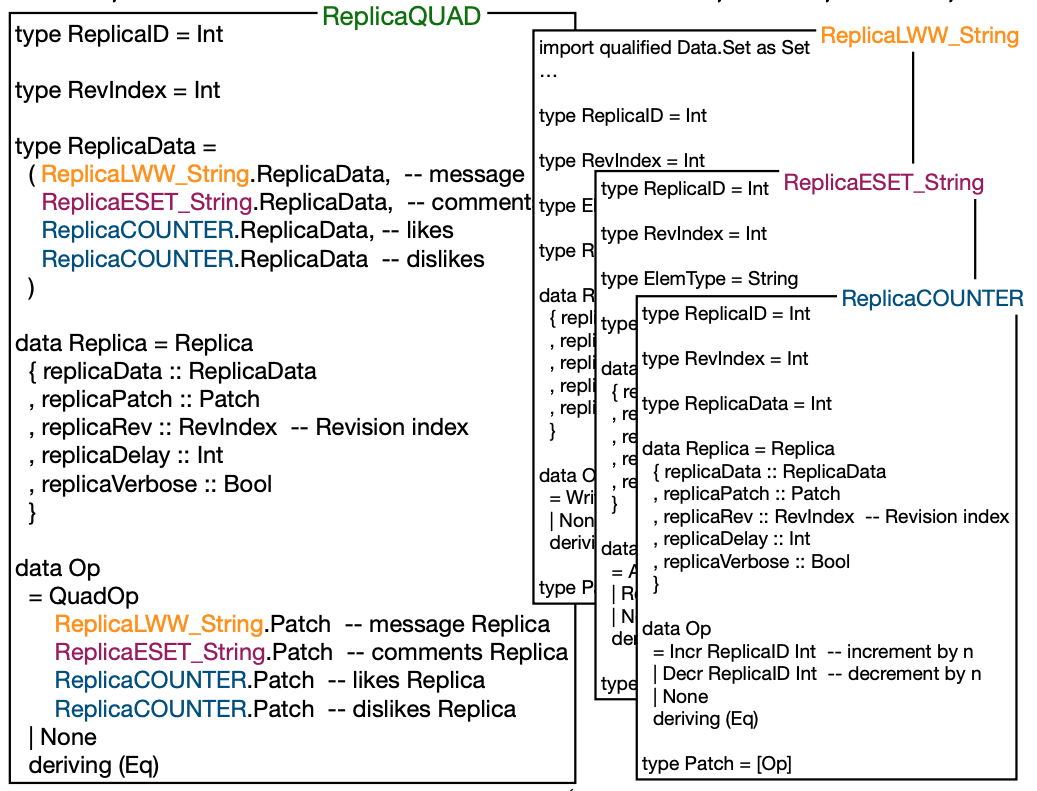}
 \end{center}
  \caption{SocialPost Quadruples - Composed Replica Data}
  \label{fig:ReplicaQUAD_Data}
\end{figure}

\begin{figure}[htb]
 \begin{center}
  \includegraphics
  [width=1.0\textwidth]{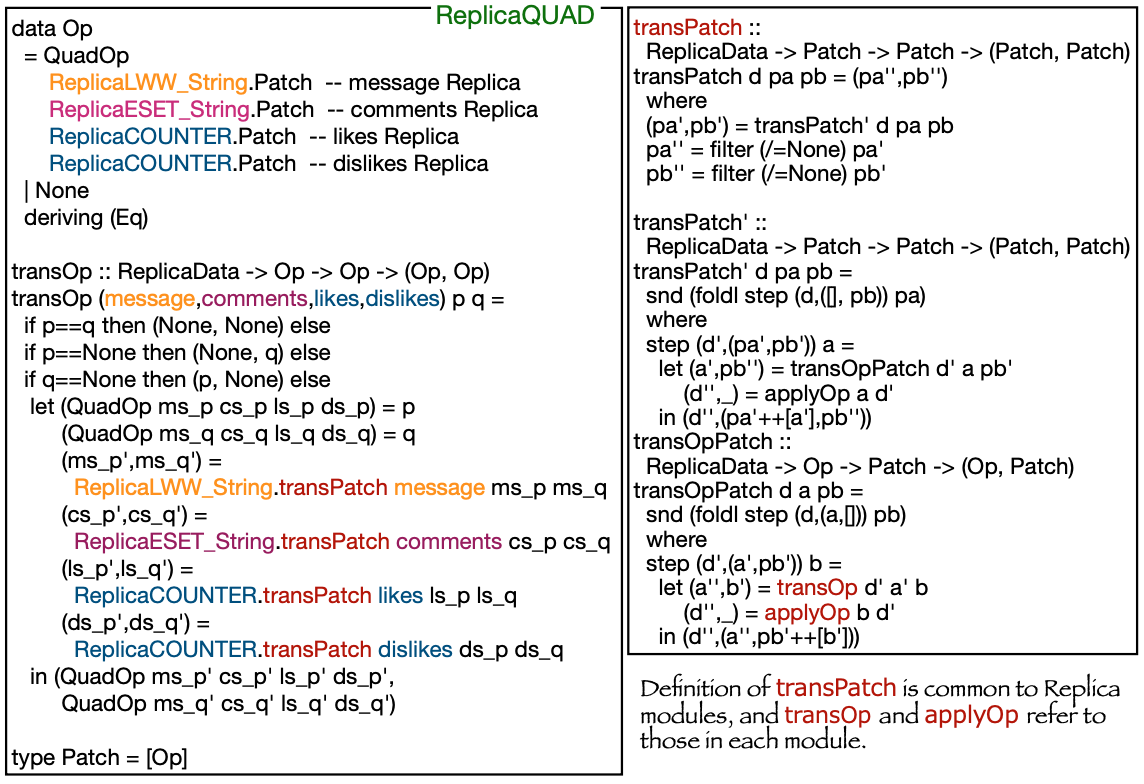}
 \end{center}
  \caption{SocialPost Quadruples - Operations and Transformation}
  \label{fig:ReplicaQUAD_Op}
\end{figure}

The $SocialPost$ quadruple
of $messages$, $comments$, $likes$, and $dislikes$
are in
Fig.\ref{fig:ReplicaQUAD_Data} and 
Fig.\ref{fig:ReplicaQUAD_Op}

\subsubsection{Map of Replica Data}
\begin{figure}[htb]
 \begin{center}
  \includegraphics
  [width=1.0\textwidth]{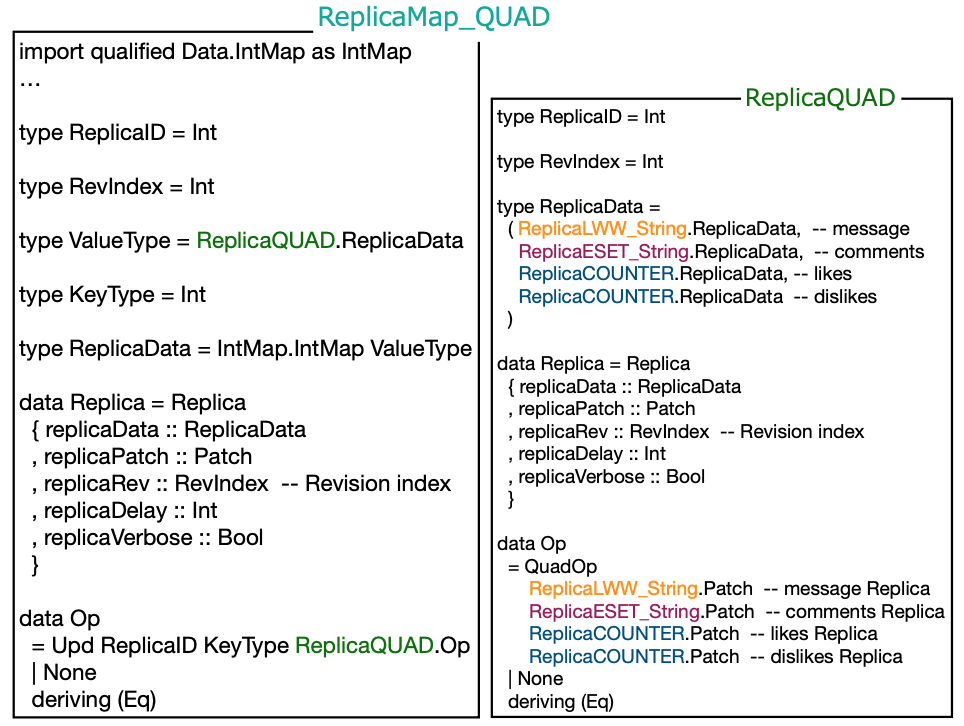}
 \end{center}
  \caption{SocialMedia Map of SocialPost Quadruples - Composed Replica Data}
  \label{fig:ReplicaMAP_QUAD_Data}
\end{figure}

\begin{figure}[htb]
 \begin{center}
  \includegraphics
  [width=1.0\textwidth]{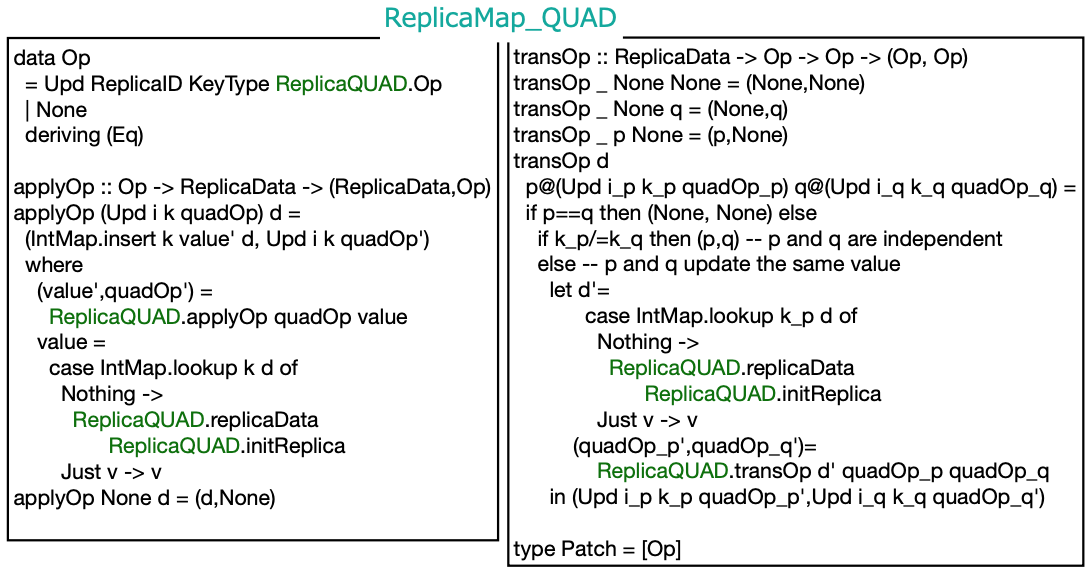}
 \end{center}
  \caption{SocialMedia Map of SocialPost Quadruples - Operations and Transformation}
  \label{fig:ReplicaMAP_QUAD_Op}
\end{figure}

The $SocialMedia$ map of the $SocialPost$ quadruples are in
Fig.\ref{fig:ReplicaMAP_QUAD_Data} and.
Fig.\ref{fig:ReplicaMAP_QUAD_Op}

\section{Remarks}
\label{sec:Remarks}
We have discussed \textit{Coordination-free Collaborative Replication} (CCR)
with several examples of construction of
composite replica types.
It may bring us the possibility of replacing the position of CRDTs with CCR, which
makes replication easier in most application areas of CRDTs.

We might have described in a top-down style that introduces
mathematical laws of the relational operation $\#_D$ for the confluence property
and then tries to find concrete representations of updating operations.
However, we have followed the process of coming up with ideas for explanation.

As mentioned in the RTCE example of CCR Agent, we have to develop some
formal method that gives the proof of TP2-Confluence property for the OT
of given updating operations.
 
Also, the cost model of CCR implementation is required for evaluating
the efficiency with benchmarking.

\newpage

\bibliographystyle{abbrv}
\bibliography{citation}



\setlength\abovecaptionskip{-10pt}

\setlength\abovecaptionskip{-10pt}

\setlength\abovecaptionskip{-10pt}

\setlength\abovecaptionskip{-10pt}

\setlength\abovecaptionskip{-10pt}

\setlength\abovecaptionskip{-10pt}

\setlength\abovecaptionskip{-10pt}

\setlength\abovecaptionskip{-10pt}

\setlength\abovecaptionskip{-10pt}

\setlength\abovecaptionskip{-10pt}

\end{document}